\documentclass[manuscript]{aastex}

\usepackage{ifpdf}

\usepackage{makeidx}

\usepackage{amssymb}

\usepackage{amsmath}

\usepackage{eurosym}

\usepackage{amsfonts}

\usepackage{rotating}

\usepackage{amssymb}

\def\c{3C~454.3 }
\def\cp{3C~454.3}

\def\mt{}
\def\es{}

\def\esc{}

\def\escc{}

\shorttitle{The exceptional gamma-ray flare of the blazar 3C
454.3} \shortauthors{Striani et al.}

\begin{document}


\title{\bf The extraordinary gamma-ray flare of the blazar 3C~454.3}

\bigskip

\author{E. Striani\altaffilmark{2,11}, S.~Vercellone\altaffilmark{17},
M.~Tavani\altaffilmark{1,2}, V. Vittorini\altaffilmark{1,2}, F.~D'Ammando\altaffilmark{1,2,17},
I.~Donnarumma\altaffilmark{1},
L.~Pacciani\altaffilmark{1},
 G.~Pucella\altaffilmark{13}, A. Bulgarelli\altaffilmark{5},
 M.~Trifoglio\altaffilmark{5}, F.~Gianotti\altaffilmark{5}, P.~Giommi\altaffilmark{14},
 A.~Argan\altaffilmark{1},
G.~Barbiellini\altaffilmark{6},
 P.~Caraveo\altaffilmark{3},
P.~W.~Cattaneo\altaffilmark{7}, A.~W.~Chen\altaffilmark{3,4},
 E.~Costa\altaffilmark{1}, G.~De
Paris\altaffilmark{1}, E. Del Monte\altaffilmark{1},
 G.~Di~Cocco\altaffilmark{5},
 Y.~Evangelista\altaffilmark{1},
M.~Feroci\altaffilmark{1}, A.~Ferrari\altaffilmark{4,18},
 M.~Fiorini\altaffilmark{3},
F.~Fuschino\altaffilmark{5}, M.~Galli\altaffilmark{8},
 A.~Giuliani\altaffilmark{3},
M.~Giusti\altaffilmark{1}, C.~Labanti\altaffilmark{5},
F.~Lazzarotto\altaffilmark{1}, P.~Lipari\altaffilmark{9},
F.~Longo\altaffilmark{6}, M.~Marisaldi\altaffilmark{5},
S.~Mereghetti\altaffilmark{3},
E.~Moretti\altaffilmark{6}, A.~Morselli\altaffilmark{11},
A.~Pellizzoni\altaffilmark{19},
F.~Perotti\altaffilmark{3}, G. Piano\altaffilmark{1,2,11},
P.~Picozza\altaffilmark{2,11}, M.~Pilia\altaffilmark{12,19},
M.~Prest\altaffilmark{12},
M.~Rapisarda\altaffilmark{13}, A.~Rappoldi\altaffilmark{7},
S. Sabatini \altaffilmark{1,11}, E.~Scalise\altaffilmark{1},
P.~Soffitta\altaffilmark{1},
 A.~Trois\altaffilmark{1},
E.~Vallazza\altaffilmark{6}, A.~Zambra\altaffilmark{3},
D.~Zanello\altaffilmark{9}, C.~Pittori\altaffilmark{14},
F.~Verrecchia\altaffilmark{14,4}, P.~Santolamazza\altaffilmark{14,4},
F.~Lucarelli\altaffilmark{14},
S.~Colafrancesco\altaffilmark{14},
L.A.~Antonelli\altaffilmark{19}, L.~Salotti\altaffilmark{15} }

\altaffiltext{1} {INAF/IASF-Roma, I-00133 Roma, Italy}
\altaffiltext{2} {Dip. di Fisica, Univ. Tor Vergata, I-00133 Roma,
Italy} \altaffiltext{3} {INAF/IASF-Milano, I-20133 Milano, Italy}
\altaffiltext{4} {CIFS-Torino, I-10133 Torino, Italy}
\altaffiltext{5} {INAF/IASF-Bologna, I-40129 Bologna, Italy}
\altaffiltext{6} {Dip. Fisica and INFN Trieste, I-34127 Trieste,
Italy} \altaffiltext{7} {INFN-Pavia, I-27100 Pavia, Italy}
\altaffiltext{8} {ENEA-Bologna, I-40129 Bologna, Italy}
\altaffiltext{9} {INFN-Roma La Sapienza, I-00185 Roma, Italy}
\altaffiltext{10} {CNR-IMIP, Roma, Italy} \altaffiltext{11} {INFN
Roma Tor Vergata, I-00133 Roma, Italy} \altaffiltext{12} {Dip. di
Fisica, Univ. Dell'Insubria, I-22100 Como, Italy}
\altaffiltext{13} {ENEA Frascati,  I-00044 Frascati (Roma), Italy}
\altaffiltext{14} {ASI Science Data Center, I-00044
Frascati (Roma), Italy} \altaffiltext{15} {Agenzia Spaziale
Italiana, I-00198 Roma, Italy} \altaffiltext{16}
{INAF-Osservatorio Astron. di Roma, Monte Porzio Catone, Italy}
\altaffiltext{17} {INAF-IASF-Palermo, via U. La Malfa 15, I-90146
Palermo, Italy}
 \altaffiltext{18} {Dip. Fisica, Universit\'a di Torino, Turin,
Italy} \altaffiltext{19} {INAF-Osservatorio Astronomico di
Cagliari, localita' Poggio dei Pini, strada 54, I-09012 Capoterra,
Italy}\altaffiltext{20} {ITAB, Via dei Vestini 33, I-66100 Chieti, Italy}





\email{EMAIL: edoardo.striani@iasf-roma.inaf.it}

\begin{abstract}

We present  the  gamma-ray data of the extraordinary flaring
activity above 100 MeV from the flat spectrum radio quasar
3C~454.3 detected by AGILE during the month of December 2009.
\cp, that has been among the most active blazars of the FSRQ type
since 2007, was detected in the gamma-ray range with a
progressively rising flux since November 10, 2009. The gamma-ray
flux reached a value comparable with that of the Vela pulsar on
December 2, 2009. Remarkably, between {\es December 2 and 3, 2009}
the source more than doubled its gamma-ray emission and became the
brightest gamma-ray source in the sky with a {\es peak flux} of
$F_{\gamma,p} = (2000 \pm 400) \times 10^{-8} \rm ph \, cm^{-2} \,
s^{-1}$ for a 1-day integration above 100 MeV. The gamma-ray
intensity decreased in the following days with the source flux
remaining at large values near $F_{\gamma} \simeq (1000 \pm 200)
\times 10^{-8} \rm ph \, cm^{-2} \, s^{-1}$ for more than a week.
This exceptional gamma-ray flare dissipated among the largest ever
detected intrinsic radiated power in gamma-rays above 100 MeV
{\esc ($L_{\gamma, source, peak} \simeq 3 \times 10^{46} \; \rm
erg \, s^{-1} $}, for a relativistic Doppler factor of $\delta
\simeq 30$). The total isotropic irradiated energy of the
month-long episode {\es in the range 100 MeV -- 3 GeV} is
$E_{\gamma,iso} \simeq 10^{56} \, \rm erg$. We report the
intensity and spectral evolution of the gamma-ray emission across
the flaring episode. We briefly discuss the important theoretical
implications of our detection.

\end{abstract}



\setcounter{footnote}{0}

\section{Introduction}


Blazars (a special class of Active Galactic Nuclei with the
relativistic jet pointing towards the Earth) show variability
across their emitted spectrum on timescales of days, months,
years. Rarely, intense gamma-ray flares are detected from blazars
with fluxes reaching values near that of the Vela pulsar (i.e.,
the brightest steady gamma-ray source in the sky with a flux of
$F_{\gamma,Vela} \simeq 900 \times 10^{-8} \rm ph \, cm^{-2} \,
s^{-1}$ above 100 MeV). Even more rarely, a blazar gamma-ray
``super-flare'' reaches intensities substantially larger than
$F_{\gamma,Vela}$, {\es as in the case of the June, 1995 flare
from PKS 1622-29 (\cite{1997ApJ...476..692M}). On these rare
occasions, the gamma-ray sky is remarkably dominated by a single
transient source.

In this \textit{Paper}, we report the observations by the AGILE
satellite of the most recent gamma-ray super-flare from the blazar
\c during the period mid-November/mid-December, 2009. During a
1-month period, this source repeatedly reached a flux near
$F_{\gamma,Vela}$ for about 2 weeks, and then produced a very
intense super-flare on {\es December 2-3}, 2009, with $F_{\gamma}
> 2 F_{\gamma,Vela} $. This flare turns out to be even more
intense than that detected from PKS 1622-29
(\cite{1997ApJ...476..692M}), and then qualifies as the most
intense gamma-ray flare ever observed from a cosmic source at
energies above 100 MeV.

\begin{figure*} 
\begin{center}
\includegraphics[width=12.cm, angle=270]{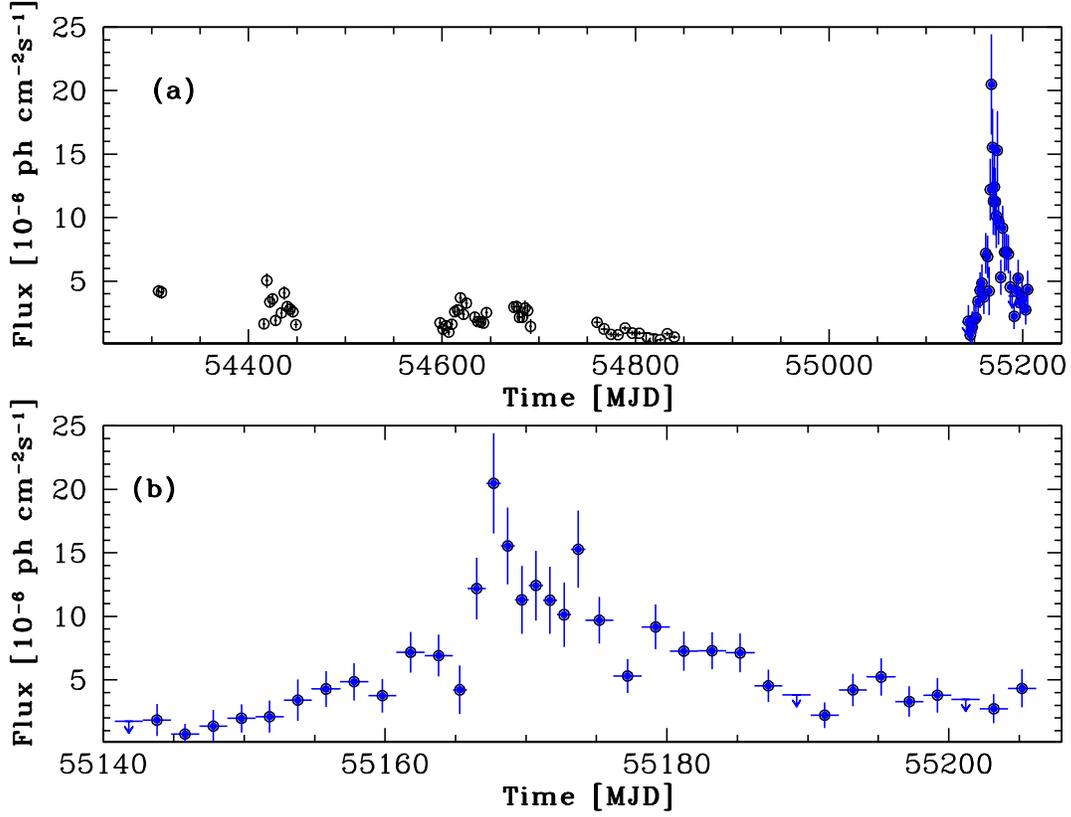}
\caption{Gamma-ray emission above 100 MeV from 3C~454.3 as
monitored by AGILE. \textit{(Upper panel:)} the 2 and 1/2 year
flux lightcurve covering the period July, 2007 - December, 2009.
The black data points are obtained with AGILE in the pointing
mode (see \cite{Vercellone2010:3C454}); the blue data points were
collected in spinning mode. \textit{(Lower panel:)} gamma-ray lightcurve obtained for the
period November 07, 2009 and January 9, 2010 (spinning mode). All
flux values are obtained with the AGILE standard maximum
likelihood analysis, using the {\es FM3.119} calibrated filter, with
standard event selection that takes into account the SAA passage
and the Earth albedo photons. The temporal bin for the blue data
points is 2 days, except for the 9-days interval centered around the peak, for which a
1-day integration was performed. {\escc Data obtained with a maximum off-axis angle $\theta_{m}=60$ degrees.}} \label{LC}
\end{center}
\end{figure*}

The blazar \c (PKS~2251$+$158; $z=0.859$) has been extensively
studied over the last two decades. A wealth of multifrequency
observations were obtained especially after the EGRET detections
above 100 MeV during the 90's in the range $F_{\gamma} = (40-140)
\times 10^{-8} \rm ph \, cm^{-2} \, s^{-1}$
(\cite{Hartman1992:3C454iauc}; \cite{Aller1997:3C454_EGRET}). The
source entered an active phase in 2000, and was very active in
2005-2006. {\esc During the May-June, 2005 period the source
showed} the strongest optical flare ever recorded in May, 2005,
{\esc and reached the}  optical magnitude  $R \simeq 12$
(\cite{vil06, Fuhrmann2006:3c454:opt}).  X-ray
(\citealt{Giommi2006:3C454_Swift}) and hard X-ray observations
(\cite{Pian2006:3C454_Integral}) {\esc covering this active phase
in 2005} detected large fluxes between 10 and 100 mCrab. \c has
been subsequently monitored very extensively at all wavelengths
\citep{vil06,vil07, 2009A&A...504L...9V,rai07,rai08a}. Starting
with the July, 2007 AGILE detection above 100 MeV
(\cite{Vercellone2008:3C454_ApJ}), \c has been very active in
gamma-rays, and certainly it can be referred as the most active
blazar during the last 2 and 1/2 years.
A series of papers describe the AGILE gamma-ray observations in 2007-2009
 showing repeated flares usually in coincidence with periods of
intense optical and X-ray enhanced activity
(\cite{Vercellone2008:3C454_ApJ},
\cite{Vercellone2008:3c454:ApJ_P1},
\cite{Vercellone2010:3C454}, \cite{2009ApJ...707.1115D});
{\esc see also the 1AGL catalog (\cite{2009A&A...506.1563P}}). The
multi-year optical evolution of \c has been presented in
\cite{2008A&A...491..755R}. Also \textit{Fermi} detected several
gamma-ray flaring episodes above 100 MeV
(\cite{2009ATel.2200....1H,2008ATel.1628....1T}), and determined
an average spectrum for the August-September, 2009 period in the
range 200 MeV -- 10 GeV showing a distinct {\es break in the
power-law spectrum at energies of 2 -- 3 GeV}
(\cite{Abdo2009:3C454}, {\es \cite{2010ApJ...710.1271A}}).


\section{AGILE and the exceptional gamma-ray flare of Dec 02, 2009}

The AGILE mission, operational  since April 2007
(\cite{2009A&A...502..995T}), is characterized by a very compact
instrument consisting of a gamma-ray imager
detector (GRID, sensitive to energy between 30 MeV and 30 GeV), and
a hard X-ray imager (Super-AGILE, sensitive in the energy range
18--60 keV). A non-imaging calorimeter (sensitive in the range
0.4--100 MeV) and an anticoincidence system complete the
instrument. The AGILE detectors are characterized by large field
of views (2.5 sr for GRID, and 1 sr for Super-AGILE), optimal
angular resolution, and good gamma-ray sensitivity especially in
the energy range 100 MeV--1 GeV.

During the period April, 2007 until end of October, 2009, AGILE
operated in a fixed-pointing mode covering about 1/5 or the entire
sky.
Due to a re-configuration of the satellite attitude control
system, in early November 2009 AGILE changed its scientific
operation mode into a ``spinning mode", with the instrument
boresight axis sweeping the sky with the angular speed of {\esc
about} 1 degree/sec. All instrument units maintained their
functionality, and in particular the GRID detector can
access about 80\% of the sky {\esc in the spinning mode}.
The observations reported in this \textit{Paper} were obtained
with AGILE operating in the spinning mode.

Because of the AGILE new pointing strategy, the FSRQ 3C~454.3 has
been constantly monitored with no gaps since early November 2009,
with a typical 2-day exposure value of $\sim 10^7 \rm \, cm^2 \,
s$ at 100 MeV.
All the gamma-ray fluxes reported in this paper were obtained with
a maximum likelihood analysis, using our {\es FM3.119} calibrated
filter. {\esc We used } a standard event selection procedure that
takes into account the SAA passage, and a standard Earth albedo
photon filtering procedure.

Fig.~\ref{LC} shows in the upper panel the {\es 2 and 1/2 year}
gamma-ray lightcurve {\esc of \c} obtained by integrating all
available AGILE gamma-ray data since July, 2007. The lower panel
of Fig.~\ref{LC} shows the detailed AGILE-GRID gamma-ray
lightcurve of  3C~454.3 obtained for the period from November 7,
2009 until January 9, 2010.
Integrating from 2009-11-10 (MJD = 55145.7) until 2009-12-02 (MJD = 55167.7),
{\esc the source shows} an increasing flux above 100 MeV from a
value of $F_{\gamma,1} \simeq$ 100 $\times \, 10^{-8}$ ph
cm$^{-2}$s$^{-1}$ up to a Vela-PSR like flux, $F_{\gamma,2}
\simeq$ 1000 $\times \, 10^{-8}$ ph cm$^{-2}$s$^{-1}$. During the
same period (November 21 and December 1-2, 2009), a gradual
optical flux increase of more than 1 magnitude in the $R$-band was
reported (\cite{2009ATel.2325....1V}), with a maximum intensity
reaching $R =14.1$ {\es on December 1-2}.

An exceptional 1-day gamma-ray emission was detected by AGILE from
\c during the time interval 2009-12-02 06:30 UT to 2009-12-03
06:30~UT (\cite{2009ATel.2326....1S}) with a peak flux of
$ F_{\gamma,p} =(2000 \pm 400) \times \, 10^{-8} \rm \, ph \, cm^{-2} \, s^{-1}$ (E $>$ 100 MeV).
During this time interval \c became the brightest gamma-ray source in the
sky with a flux above 100 MeV $F_{\gamma,p} > 2 \,
F_{\gamma,Vela}$. This flux
exceeded that one of the previous day, showing a rapid increase
(about 80\%) within 24 hours. {\esc
Integrating from 2009-11-29 19:00 until 2009-12-01 17:00 (MJD
55164.8 -- 55166.7) we obtain the 2-day averaged gamma-ray flux
$F'_{\gamma}(2-day) = (688\pm 160) \times \, 10^{-8} \rm \, ph \,
cm^{-2} \, s^{-1}$. Integrating from 2009-12-01 17:00 until
2009-12-03 17:00 (MJD 55166.7 -- 55168.7) we obtain the flux $
F_{\gamma,p} (2-day) = (1680 \pm 240) \times \, 10^{-8} \rm \, ph
\, cm^{-2} \, s^{-1} $. The statistical significance of the 2-day
averaged flux variability turns out to be above 3 $\sigma$.}
{\esc Following the super-flare episode,} the source flux
decreased to an average value near 1000 $\times 10^{-8}$ ph
cm$^{-2}$s$^{-1}$ (E$>$ 100 MeV) during the following 10 days, and
later decreased to an average flux of 400 $\times 10^{-8}$ ph
cm$^{-2}$s$^{-1}$ (E$>$ 100 MeV).



Multifrequency observations in the IR, optical, X-ray and
$\gamma$-ray bands have been reported for this extraordinary
gamma-ray flaring activity including those by \textit{Fermi}/LAT
(\cite{2009ATel.2328....1E}), \textit{Swift}/XRT
(\cite{2009ATel.2329....1S}), \textit{Swift}/BAT
(\cite{2009ATel.2330....1K}), SMARTS/ANDICAM
(\cite{2009ATel.2332....1B}), and the Kanata telescope (\cite
{2009ATel.2333....1S}), and during the following days by
\textit{INTEGRAL}/IBIS (\cite{2009ATel.2344....1V}, MIRO
(\cite{2009ATel.2345....1B}) and  ARIES
(\cite{2009ATel.2352....1G}).

In order to study the gamma-ray emission in the AGILE main
spectral bands, we plot in Fig.~\ref{HR} the 1-day integrated
lightcurves in two energy ranges, a ``soft" band with $100 \, {\rm
MeV} \, < \, E \, < \, 400$~MeV (blue curve), and  a ``hard" band
with {\es $400 \, {\rm
MeV} \, < \, E \, < \, 3$~GeV} (red curve).
{\escc The shape of the lightcurves in the two energy bands shows a possible hint
that the source emission hardens across the super-flare episode. Fig.~\ref{HR} shows, indeed,
a substantial increase of the ``hard" flux, while a flattening is present in the ``soft" lightcurve.
The full band lightcurve reported in Fig.~\ref{LC} seems to be more influenced by the behavior of the
hard band emission across the super-flare episode.}


\begin{figure} 
\begin{center}
\includegraphics[width=15.cm]{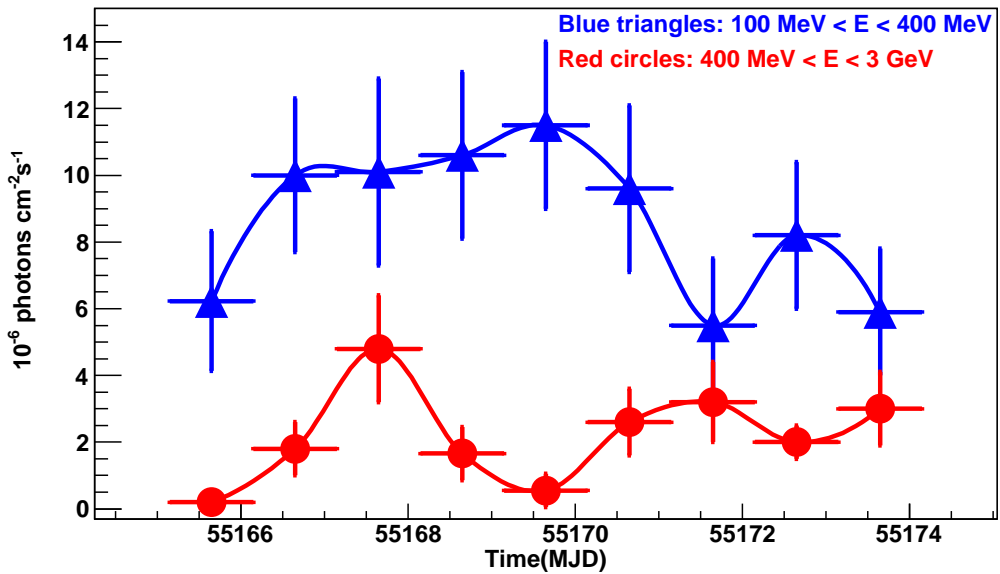}
\caption{The 1-day gamma-ray lightcurve of the 3C454.3 between
{\es November 30 and December 8}, 2009 in different spectral bands.
 \textit{(Blue curve and data points)}: lightcurve for photon energies
100 MeV $ \, < \, E \, <$~400~MeV.   \textit{(Red curve and data
points)}; lightcurve for photon energies {\es $400 \, {\rm
MeV} \, < \, E \, < \, 3$~GeV}. The
super-flare episode {\escc occurred on} 2-3 December, 2009 ({\es MJD = 55167.7}).
{\escc Data obtained with a maximum off-axis angle $\theta_{m}=40$ degrees.}} \label{HR}
\end{center}
\end{figure}

\begin{figure}[h!]
\begin{center}
\noindent
\includegraphics[width=15.cm]{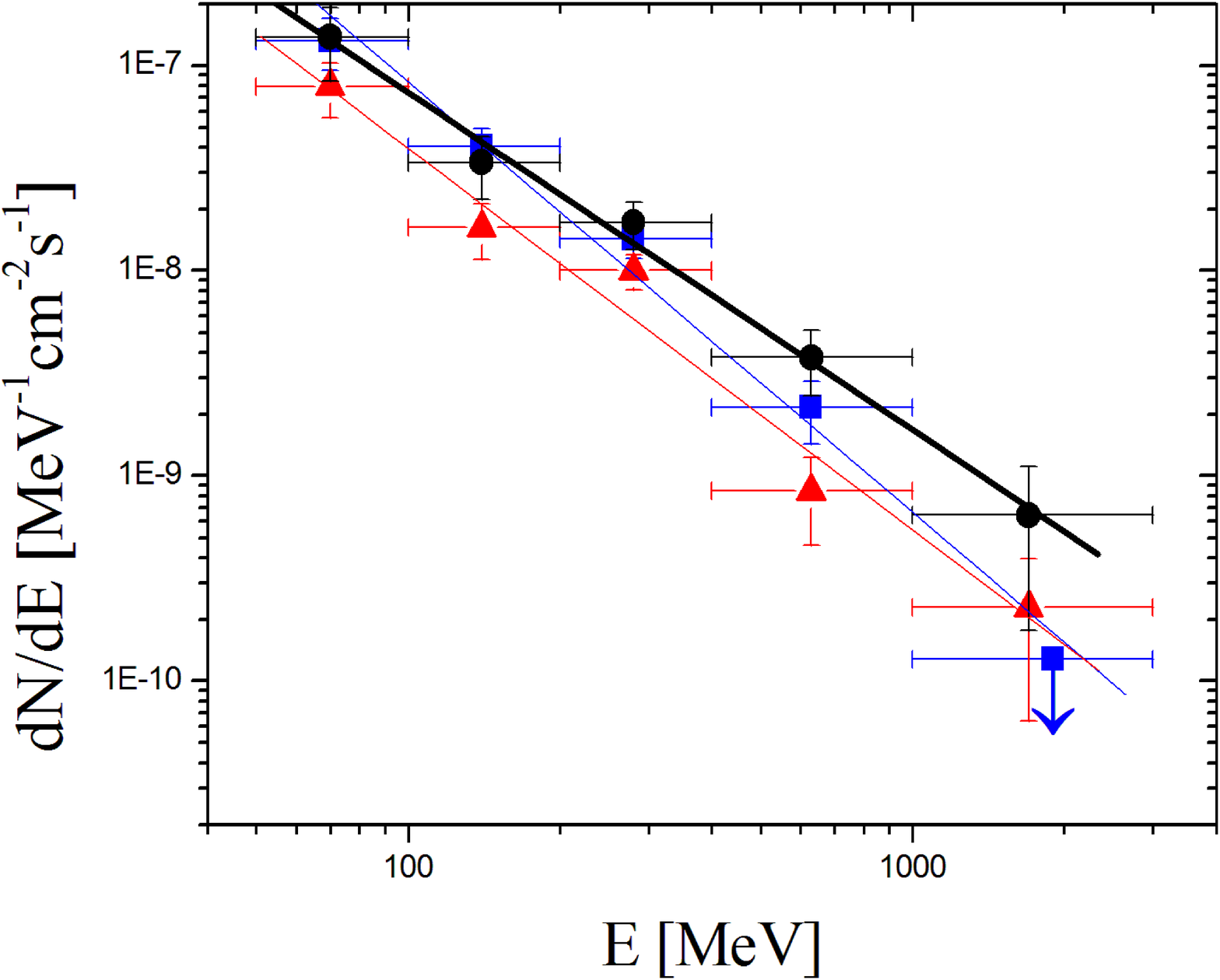}
\caption{
Photon number differential energy spectra of 3C~454.3 before,
during and following the Dec. 2-3, 2009 super-flare. \textit{Red triangles:} the {\escc 25-day} spectrum integrated for the period
between 2009-11-06 05:00 UT and 2009-12-01 01:00 UT. \textit{Black circles:} the {\escc 2-day} spectrum integrated for the period
between 2009-12-01 12:00 UT and 2009-12-03 17:00 UT. \textit{Blue squares:} the {\escc 3-day} spectrum integrated for the period between
2009-12-03 17:00 UT and 2009-12-06 16:00 UT. {\esc Solid lines
show the spectral slopes discussed in the main text}. {\escc Data obtained with a maximum off-axis angle $\theta_{m}=40$ degrees.}}
\label{spectra}
\end{center}
\end{figure}

We carried out a {\es time-resolved} spectral analysis of 3C~454.3
dividing the period of exceptional gamma-ray activity in three
time intervals. Fig.~\ref{spectra} shows our results for: (1) a
25-day integrated period before the super-flare (interval-1), (2)
the super-flare episode integrated over 2 days (interval-2), and (3)
the following 3 days (interval-3). {\es The 3- and 25-day
integrations are chosen to provide a good statistical sample.}

{\escc The spectral behavior during these three intervals seems to confirm the
hardening deduced by the ``soft"  and ``hard"  gamma-ray properties of
the source emission during the super-flare. The super-flare spectrum is well fitted
between 100 MeV - 1 GeV by a single power-law with {\esc photon} index $\alpha = 1.66 \,
\pm \,0.32$ (statistical error only, {\esc defined as
$dN_{\gamma}/dE \propto E^{-\alpha}$}). A \textit{single power-law
fit} for the pre- and post-flare spectra gives $\alpha = 1.85 \,
\pm \,0.26$ {\esc (interval-1)} and $\alpha = 2.04 \, \pm \, 0.26$
{\esc (interval-3)}, respectively\footnote{However, it has to be noticed that these values are all consistent
within 1-sigma level}. The 
pre- and post-super-flare spectra show a curvature with a peak
energy  $E_p$ of the $\nu F_{\nu}$ spectrum, $E_p \simeq
300$~MeV.}

%
%


We {\esc also} performed a {\esc refined} temporal analysis of the
super-flare episode, and studied the gamma-ray lightcurve with
temporal bins of different durations (24 hrs, as well as 12 - 6 -
3 hrs). We extracted gamma-ray photons from a radius of $2^o$ for
the period 55165-55173~MJD, and obtained the corresponding
lightcurves with different binning.
Restricting {\es the analysis} to the super-flare period
(55166-55169~MJD) and to the 6-hr bin lightcurve above 100 MeV, we
detect a 3 $\sigma$ peak (above a 2-day average) during the {\esc
6-hour} period 2009-12-02 10:30~UT and 2009-12-02 16:30~UT. This
sharp increase (for AGILE due essentially to photons above 400
MeV) is in temporal agreement with the 6-hr increase also reported
by \textit{Fermi} (\cite{2009ATel.2328....1E}).

\section{Discussion}

\c turns out to be the {\es brightest} and most active gamma-ray
blazar detected above 100 MeV since the beginning of operations of
the new generation gamma-ray instruments (AGILE and
\textit{Fermi}). Many gamma-ray flares have been detected from
this source {\es in the last two years}, and the most recent
flaring at the end of the year 2009 is the culmination of a
very active phase. The
gamma-ray super-flare of early December, 2009 is remarkable for
many reasons.

\textit{(1)} It reaches the strongest ever gamma-ray flux detected
from a blazar, with an apparent (isotropic) 1-day peak
luminosity of $L_{\gamma, iso,p} \simeq 6 \times
10^{49} \, \rm erg \, s^{-1}$ above 100 MeV.
{\escc We note that for a \c black hole mass of
$M \simeq 2 \times 10^9 \, M_{\odot}$ (\cite{2002ApJ...579..530W}) the
observed isotropic  gamma-ray luminosity is apparently strongly
super-Eddington. However, taking into account the radiation
pressure of a fraction of spherical surface for a jet opening
angle $\phi \sim 5^o$ (\cite{2004ApJ...615L..81G}), we obtain an
observed radiated luminosity of order of 1/10 of the  Eddington
limit obtained for spherical accretion ($L_E \simeq 3 \times
10^{47} \rm \, erg \, s^{-1}$). For an efficiency near 10\% of
kinetic energy conversion into gamma-ray radiation and maximal
extraction of accretion power into jet kinetic power, the deduced
accretion rate can reach the Eddington limit during the
super-flare episode.}
{\esc By rescaling {\escc the peak luminosity}  value with the relativistic beaming and Doppler factors,
we obtain {\escc the intrinsic peak luminosity} $L_{\gamma, source,p} = L_{\gamma, iso,p} \, \Gamma^2 \,
\delta^{-4}$ (e.g., \cite{2003ApJ...593..667M}). There are uncertainties for the value of the Doppler
factor $\delta = \Gamma^{-1} \, (1 - \beta \, \cos \theta)^{-1}$
(with $\Gamma= (1-\beta^2)^{-1/2} $, and $\beta$ the bulk jet
velocity), for which values between 20 and 40 are found in
literature. We adopt here, for consistency with other {\es recent}
investigations (e.g., \cite{Vercellone2010:3C454}), the value
$\delta \simeq 30 $ (see also \cite {2009arXiv0911.4924S}). The
corresponding values of the bulk Lorentz factor and the angle
$\theta$ between the jet axis and line of sight are in the
approximate ranges $15 < \Gamma <20 $, and $1^o < \theta < 3^o $.
We adopt the values $\Gamma = 20$, and $\theta=1.2^o$, and  obtain
$ L_{\gamma, source,p} \simeq 3 \times 10^{46} \, \rm erg \,
s^{-1}$.}

 \textit{(2)} The peak
gamma-ray emission is characterized by a very short risetime (6-12
hrs), {\escc and shows a spectral evolution with a hard
spectral component that modifies the pre- and
post-peak spectra} (see Figures 2 and 3).

\textit{(3)} The total isotropic irradiated energy in the range
100 MeV -- 3 GeV during the 2-month period 55146-55205~MJD is
$E_{\gamma,iso} \simeq 10^{56} \, \rm erg$. This implies an
intrinsic total radiated energy above 100 MeV of
$E_{\gamma,source} = E_{\gamma,iso} \, \Gamma^2 \, \delta^{-4}
\simeq  5 \times 10^{52} \rm \, erg \simeq 1/40 \, M_{\odot}$. For
comparison, the total energy irradiated in the gamma-ray band by
the PKS 1622-29 {\es(at z $\simeq$ 0.8, assuming the same
parameters as for \cp) during the flare in 1995 was}
$E_{\gamma,source} \simeq 3 \times 10^{52} \rm \, erg$.

The \c super-flare phenomenon is intrinsically broad-band in
nature, and a satisfactory picture of the emission mechanism can
be obtained only from a complete multifrequency account of the spectral
evolution (see \cite{2010arXiv1005.3263P}). We briefly focus here
on the gamma-ray spectral features as shown in Figs. 2 and 3.
Among the possible mechanisms that can account for {escc a}
spectral hardening and subsequent decay observed during and
following the \c super-flare, we mention here
{\mt one possibility:  the } injection of energetic electrons with
an energy cutoff larger by a factor of $\sim 3$ than the
$\gamma_{c}$ applicable to the pre- and post-flare conditions.
This extra-acceleration, that has to occur with a comoving
timescale less than $\tau_a \simeq \rm (1 \, day) \, \delta \, (1
+ z)^{-1}$, influences the whole SED modifying the synchrotron,
{\es synchrotron self-Compton (SSC)} and external inverse Compton
components.
The rapid spectral variation inferred from Fig.~3 {\mt indeed}
argues for substantial cooling of the particle distribution
function.
{\mt The } additional component {\mt of } energized particles
{\mt may be}  the manifestation of drastic modification of the
inner parts of the disk/jet system that produces the rapidly
variable high-energy emission. Previous broad-band {\mt SED}
determinations of \c  for the gamma-ray activity detected by AGILE
in 2007, 2008 and 2009 (\cite{Vercellone2008:3C454_ApJ},
\cite{Vercellone2008:3c454:ApJ_P1}, \cite{2009ApJ...707.1115D},
{\mt see also Raiteri et al. 2007 for the detection of little and
big blue bumps in the optical spectra}) constitute {\mt an}
unprecedentedly constraining database from which further
theoretical modelling can be developed.

\section{Conclusions}

\c reveals itself as the most prolific gamma-ray blazar during the
last 3 years, and is
dominating the {\esc gamma-ray} sky above 100 MeV since mid-2007.
During the period of December, 2009 the source showed a dramatic
activity reaching and maintaining for several weeks a
flux above 100 MeV comparable or larger than the brightest
persistent gamma-ray source such as the Vela pulsar. During the
period Dec. 2-3, 2009 \c produced a
super-flare that turns out to be the brightest blazar emission
episode above 100 MeV ever detected. The AGILE satellite followed
{\esc in a continuous way} the daily evolution of the flaring
activity of \cp.
Even though a comprehensive picture of the physical mechanism at
work
can be obtained only from a
multifrequency collection of simultaneous data, restricting
ourselves to the gamma-ray range in any case provides very
important information on the physics of the source. For a detailed
theoretical modelling and a broad spectral evolution of the \c
exceptional activity see \cite{2010arXiv1005.3263P}.

\section{Acknowledgements}

We thank an anonymous referee for his/her comments that improved
our paper.
 The AGILE mission is funded by the Italian Space Agency
 with scientific and programmatic
participation by the Italian Institute of Astrophysics and the
Italian Institute of Nuclear Physics. {\mt Research partially
funded through the ASI contract n. I/089/06/2.}

\clearpage
\bibliographystyle{apj}


\clearpage

\begin{thebibliography}{37}
\expandafter\ifx\csname natexlab\endcsname\relax\def\natexlab#1{#1}\fi

\bibitem[{{Abdo} {et~al.}(2009){Abdo}, {Ackermann}, {Ajello}, {Atwood},
  {Axelsson}, {Baldini}, {Ballet}, {Barbiellini}, {Bastieri}, {Battelino},
  {Baughman}, {Bechtol}, {Bellazzini}, {Berenji}, {Blandford}, {Bloom},
  {Bonamente}, {Borgland}, {Bouvier}, {Bregeon}, {Brez}, {Brigida}, {Bruel},
  {Burnett}, {Caliandro}, {Cameron}, {Caraveo}, {Casandjian}, {Cavazzuti},
  {Cecchi}, {Charles}, {Chaty}, {Chekhtman}, {Cheung}, {Chiang}, {Ciprini},
  {Claus}, {Cohen-Tanugi}, {Cominsky}, {Conrad}, {Costamante}, {Cutini},
  {Dermer}, {de Angelis}, {de Palma}, {Digel}, {Silva}, {Donato}, {Drell},
  {Dubois}, {Dumora}, {Farnier}, {Favuzzi}, {Focke}, {Foschini}, {Frailis},
  {Fuhrmann}, {Fukazawa}, {Funk}, {Fusco}, {Gargano}, {Gasparrini}, {Gehrels},
  {Germani}, {Giebels}, {Giglietto}, {Giommi}, {Giordano}, {Glanzman},
  {Godfrey}, {Grenier}, {Grondin}, {Grove}, {Guillemot}, {Guiriec}, {Hanabata},
  {Harding}, {Hartman}, {Hayashida}, {Hays}, {Hughes}, {J{\'o}hannesson},
  {Johnson}, {Johnson}, {Johnson}, {Kamae}, {Katagiri}, {Kataoka}, {Kawai},
  {Kerr}, {Kn{\"o}dlseder}, {Kocian}, {Kuehn}, {Kuss}, {Latronico}, {Lee},
  {Lemoine-Goumard}, {Longo}, {Loparco}, {Lott}, {Lovellette}, {Lubrano},
  {Madejski}, {Makeev}, {Massaro}, {Mazziotta}, {McEnery}, {McGlynn}, {Meurer},
  {Michelson}, {Mitthumsiri}, {Mizuno}, {Moiseev}, {Monte}, {Monzani},
  {Morselli}, {Moskalenko}, {Murgia}, {Nolan}, {Norris}, {Nuss}, {Ohsugi},
  {Omodei}, {Orlando}, {Ormes}, {Paneque}, {Panetta}, {Parent}, {Pelassa},
  {Pepe}, {Pesce-Rollins}, {Piron}, {Porter}, {Rain{\`o}}, {Rando}, {Razzano},
  {Reimer}, {Reimer}, {Reposeur}, {Reyes}, {Ritz}, {Rochester}, {Rodriguez},
  {Rahoui}, {Ryde}, {Sadrozinski}, {Sambruna}, {Sanchez}, {Sander},
  {Parkinson}, {Sgr{\`o}}, {Shaw}, {Smith}, {Smith}, {Spandre}, {Spinelli},
  {Starck}, {Strickman}, {Suson}, {Tajima}, {Takahashi}, {Takahashi}, {Tanaka},
  {Thayer}, {Thayer}, {Thompson}, {Tibaldo}, {Torres}, {Tosti}, {Tramacere},
  {Uchiyama}, {Usher}, {Vilchez}, {Villata}, {Vitale}, {Waite}, {Winer},
  {Wood}, {Ylinen}, {Zensus}, \& {Ziegler}}]{Abdo2009:3C454}
{Abdo}, A.~A. {et~al.} 2009, \apj, 699, 817

\bibitem[{{Abdo} {et~al.}(2010{\natexlab{b}}){Abdo}, {Ackermann}, {Ajello},
  {Atwood}, {Axelsson}, {Baldini}, {Ballet}, {Barbiellini}, {Bastieri},
  {Bechtol}, {Bellazzini}, {Berenji}, {Blandford}, {Bloom}, {Bonamente},
  {Borgland}, {Bouvier}, {Bregeon}, {Brez}, {Brigida}, {Bruel}, {Burnett},
  {Buson}, {Caliandro}, {Cameron}, {Caraveo}, {Carrigan}, {Casandjian},
  {Cavazzuti}, {Cecchi}, {{\c C}elik}, {Charles}, {Chekhtman}, {Cheung},
  {Chiang}, {Ciprini}, {Claus}, {Cohen-Tanugi}, {Conrad}, {Cutini}, {Dermer},
  {de Angelis}, {de Palma}, {Digel}, {Silva}, {Drell}, {Dubois}, {Dumora},
  {Farnier}, {Favuzzi}, {Fegan}, {Focke}, {Fortin}, {Frailis}, {Fukazawa},
  {Funk}, {Fusco}, {Gargano}, {Gasparrini}, {Gehrels}, {Germani}, {Giebels},
  {Giglietto}, {Giommi}, {Giordano}, {Glanzman}, {Godfrey}, {Grenier},
  {Grondin}, {Grove}, {Guillemot}, {Guiriec}, {Harding}, {Hartman},
  {Hayashida}, {Hays}, {Healey}, {Horan}, {Hughes}, {Jackson},
  {J{\'o}hannesson}, {Johnson}, {Johnson}, {Kamae}, {Katagiri}, {Kataoka},
  {Kawai}, {Kerr}, {Kn{\"o}dlseder}, {Kuss}, {Lande}, {Latronico},
  {Lemoine-Goumard}, {Longo}, {Loparco}, {Lott}, {Lovellette}, {Lubrano},
  {Madejski}, {Makeev}, {Mazziotta}, {McConville}, {McEnery}, {Meurer},
  {Michelson}, {Mitthumsiri}, {Mizuno}, {Moiseev}, {Monte}, {Monzani},
  {Morselli}, {Moskalenko}, {Murgia}, {Nolan}, {Norris}, {Nuss}, {Ohsugi},
  {Omodei}, {Orlando}, {Ormes}, {Paneque}, {Panetta}, {Parent}, {Pelassa},
  {Pepe}, {Persic}, {Pesce-Rollins}, {Piron}, {Porter}, {Rain{\`o}}, {Rando},
  {Razzano}, {Reimer}, {Reimer}, {Reposeur}, {Ritz}, {Rochester}, {Rodriguez},
  {Romani}, {Roth}, {Ryde}, {Sadrozinski}, {Sanchez}, {Sander}, {Saz
  Parkinson}, {Scargle}, {Sgr{\`o}}, {Siskind}, {Smith}, {Smith}, {Spandre},
  {Spinelli}, {Strickman}, {Suson}, {Tajima}, {Takahashi}, {Takahashi},
  {Tanaka}, {Thayer}, {Thayer}, {Thompson}, {Tibaldo}, {Torres}, {Tosti},
  {Tramacere}, {Uchiyama}, {Usher}, {Vasileiou}, {Vilchez}, {Villata},
  {Vitale}, {Waite}, {Wang}, {Winer}, {Wood}, {Ylinen}, \&
  {Ziegler}}]{2010ApJ...710.1271A}
---. 2010{\natexlab{b}}, \apj, 710, 1271


\bibitem[{{Aller} {et~al.}(1997){Aller}, {Marscher}, {Hartman}, {Aller},
  {Aller}, {Balonek}, {Begelman}, {Chiaberge}, {Clements}, {Collmar}, {de
  Francesco}, {Gear}, {Georganopoulos}, {Ghisellini}, {Glass},
  {Gonzalez-Perez}, {Heinimaki}, {Herter}, {Hooper}, {Hughes}, {Johnson},
  {Katajainen}, {Kidger}, {Kraus}, {Lanteri}, {Lawrence}, {Lichti}, {Lin},
  {Madejski}, {McNaron-Brown}, {Moore}, {Mukherjee}, {Nair}, {Nilsson},
  {Peila}, {Pierkowski}, {Pohl}, {Pursimo}, {Raiteri}, {Reich}, {Robson},
  {Sillanpaa}, {Sikora}, {Smith}, {Steppe}, {Stevens}, {Takalo}, {Terasranta},
  {Tornikoski}, {Valtaoja}, {von Montigny}, {Villata}, {Wagner}, {Wichmann}, \&
  {Witzel}}]{Aller1997:3C454_EGRET}
{Aller}, M.~F. {et~al.} 1997, in American Institute of Physics Conference
  Series, Vol. 410, Proceedings of the Fourth Compton Symposium, ed. C.~D.
  {Dermer}, M.~S. {Strickman}, \& J.~D. {Kurfess}, 1423

\bibitem[{{Baliyan} {et~al.}(2009){Baliyan}, {Ganesh}, {Chandra}, \&
  {Joshi}}]{2009ATel.2345....1B}
{Baliyan}, K., {Ganesh}, S., {Chandra}, S., \& {Joshi}, U. 2009, The
  Astronomer's Telegram, 2345, 1

\bibitem[{{Bonning} {et~al.}(2009){Bonning}, {Bailyn}, {Buxton}, {Chatterjee},
  {Coppi}, {Isler}, {Urry}, {Maraschi}, \& {Fossati}}]{2009ATel.2332....1B}
{Bonning}, E. {et~al.} 2009, The Astronomer's Telegram, 2332, 1

\bibitem[{{Donnarumma} {et~al.}(2009){Donnarumma}, {Pucella}, {Vittorini},
  {D'Ammando}, {Vercellone}, {Raiteri}, {Villata}, {Perri}, {Chen}, {Smart},
  {Kataoka}, {Kawai}, {Mori}, {Tosti}, {Impiombato}, {Takahashi}, {Sato},
  {Tavani}, {Bulgarelli}, {Chen}, {Giuliani}, {Longo}, {Pacciani}, {Argan},
  {Barbiellini}, {Boffelli}, {Caraveo}, {Cattaneo}, {Cocco}, {Contessi},
  {Costa}, {Del Monte}, {De Paris}, {Di Cocco}, {Evangelista}, {Feroci},
  {Ferrari}, {Fiorini}, {Froysland}, {Frutti}, {Fuschino}, {Galli}, {Gianotti},
  {Labanti}, {Lapshov}, {Lazzarotto}, {Lipari}, {Marisaldi}, {Mastropietro},
  {Mereghetti}, {Morelli}, {Moretti}, {Morselli}, {Pellizzoni}, {Perotti},
  {Piano}, {Picozza}, {Pilia}, {Porrovecchio}, {Prest}, {Rapisarda},
  {Rappoldi}, {Rubini}, {Sabatini}, {Scalise}, {Soffitta}, {Striani},
  {Trifoglio}, {Trois}, {Vallazza}, {Zambra}, {Zanello}, {Pittori},
  {Santolamazza}, {Verrecchia}, {Giommi}, {Antonelli}, {Colafrancesco}, \&
  {Salotti}}]{2009ApJ...707.1115D}
{Donnarumma}, I. {et~al.} 2009, \apj, 707, 1115

\bibitem[{{Escande} \& {Tanaka}(2009)}]{2009ATel.2328....1E}
{Escande}, L., \& {Tanaka}, Y.~T. 2009, The Astronomer's Telegram, 2328, 1

\bibitem[{{Fuhrmann} {et~al.}(2006){Fuhrmann}, {Cucchiara}, {Marchili},
  {Tosti}, {Nucciarelli}, {Ciprini}, {Molinari}, {Chincarini}, {Zerbi},
  {Covino}, {Pian}, {Meurs}, {Testa}, {Vitali}, {Antonelli}, {Conconi},
  {Cutispoto}, {Malaspina}, {Nicastro}, {Palazzi}, \&
  {Ward}}]{Fuhrmann2006:3c454:opt}
{Fuhrmann}, L. {et~al.} 2006, \aap, 445, L1

\bibitem[{{Giommi} {et~al.}(2006){Giommi}, {Blustin}, {Capalbi},
  {Colafrancesco}, {Cucchiara}, {Fuhrmann}, {Krimm}, {Marchili}, {Massaro},
  {Perri}, {Tagliaferri}, {Tosti}, {Tramacere}, {Burrows}, {Chincarini},
  {Falcone}, {Gehrels}, {Kennea}, \& {Sambruna}}]{Giommi2006:3C454_Swift}
{Giommi}, P. {et~al.} 2006, \aap, 456, 911

\bibitem[{{Gopal-Krishna} {et~al.}(2004){Gopal-Krishna}, {Dhurde}, \&
  {Wiita}}]{2004ApJ...615L..81G}
{Gopal-Krishna}, {Dhurde}, S., \& {Wiita}, P.~J. 2004, \apjl, 615, L81


\bibitem[{{Gupta} {et~al.}(2009){Gupta}, {Gaur}, \&
  {Rani}}]{2009ATel.2352....1G}
{Gupta}, A.~C., {Gaur}, H., \& {Rani}, B. 2009, The Astronomer's Telegram,
  2352, 1

\bibitem[{{Hartman} {et~al.}(1992){Hartman}, {Bertsch}, {Fichtel}, {Hunter},
  {Kwok}, {Mattox}, {Sreekumar}, {Thompson}, {Kniffen}, {Lin}, {Michelson},
  {Nolan}, {Schneid}, {Kanbach}, {Mayer-Hasselwander}, {von Montigny},
  {Pinkau}, {Rothermel}, \& {Sommer}}]{Hartman1992:3C454iauc}
{Hartman}, R.~C. {et~al.} 1992, \iaucirc, 5477, 2

\bibitem[{{Hill}(2009)}]{2009ATel.2200....1H}
{Hill}, A.~B. 2009, The Astronomer's Telegram, 2200, 1

\bibitem[{{Krimm} {et~al.}(2009){Krimm}, {Barthelmy}, {Baumgartner},
  {Cummings}, {Fenimore}, {Gehrels}, {Kadler}, {Markwardt}, {Palmer},
  {Sakamoto}, {Skinner}, {Stamatikos}, {Tueller}, \&
  {Ukwatta}}]{2009ATel.2330....1K}
{Krimm}, H.~A. {et~al.} 2009, The Astronomer's Telegram, 2330, 1

\bibitem[{{Maraschi} \& {Tavecchio}(2003)}]{2003ApJ...593..667M}
{Maraschi}, L., \& {Tavecchio}, F. 2003, \apj, 593, 667

\bibitem[{{Mattox} {et~al.}(1997){Mattox}, {Wagner}, {Malkan}, {McGlynn},
  {Schachter}, {Grove}, {Johnson}, \& {Kurfess}}]{1997ApJ...476..692M}
{Mattox}, J.~R., {Wagner}, S.~J., {Malkan}, M., {McGlynn}, T.~A., {Schachter},
  J.~F., {Grove}, J.~E., {Johnson}, W.~N., \& {Kurfess}, J.~D. 1997, \apj, 476,
  692

\bibitem[{{Pacciani} {et~al.}(2010){Pacciani}, {Vittorini}, {Tavani},
  {Fiocchi}, {Vercellone}, {D'Ammando}, {Sakamoto}, {Pian}, {Raiteri},
  {Villata}, {Sasada}, {Itoh}, {Yamanaka}, {Uemura}, {Striani}, {Fugazza},
  {Tiengo}, {Krimm}, {Stroh}, {Falcone}, {Curran}, {Sadun}, {Lahteenmaki},
  {Tornikoski}, {Aller}, {Aller}, {Lin}, {Larionov}, {Leto}, {Takalo},
  {Berdyugin}, {Gurwell}, {Bulgarelli}, {Chen}, {Donnarumma}, {Giuliani},
  {Longo}, {Pucella}, {Argan}, {Caraveo}, {Cattaneo}, {Cocco}, {Costa}, {De
  Paris}, {Del Monte}, {Di Cocco}, {Evangelista}, {Ferrari}, {Feroci},
  {Fiorini}, {Fuschino}, {Galli}, {Gianotti}, {Labanti}, {Lapshov},
  {Lazzarotto}, {Lipari}, {Marisaldi}, {Mereghetti}, {Morelli}, {Moretti},
  {Morselli}, {Pellizzoni}, {Perotti}, {Piano}, {Picozza}, {Pilia}, {Prest},
  {Rapisarda}, {Rappoldi}, {Rubini}, {Sabatini}, {Soffitta}, {Trifoglio},
  {Trois}, {Vallazza}, {Zanello}, {Colafrancesco}, {Pittori}, {Verrecchia},
  {Santolamazza}, {Lucarelli}, {Giommi}, \& {Salotti}}]{2010arXiv1005.3263P}
{Pacciani}, L. {et~al.} 2010, ArXiv e-prints, ApJ Letters accepted


\bibitem[{{Pian} {et~al.}(2006){Pian}, {Foschini}, {Beckmann}, {Soldi},
  {T{\"u}rler}, {Gehrels}, {Ghisellini}, {Giommi}, {Maraschi}, {Pursimo},
  {Raiteri}, {Tagliaferri}, {Tornikoski}, {Tosti}, {Treves}, {Villata}, {Barr},
  {Courvoisier}, {di Cocco}, {Hudec}, {Fuhrmann}, {Malaguti}, {Persic},
  {Tavecchio}, \& {Walter}}]{Pian2006:3C454_Integral}
{Pian}, E. {et~al.} 2006, \aap, 449, L21

\bibitem[{{Pittori} {et~al.}(2009){Pittori}, {Verrecchia}, {Chen},
  {Bulgarelli}, {Pellizzoni}, {Giuliani}, {Vercellone}, {Longo}, {Tavani},
  {Giommi}, {Barbiellini}, {Trifoglio}, {Gianotti}, {Argan}, {Antonelli},
  {Boffelli}, {Caraveo}, {Cattaneo}, {Cocco}, {Colafrancesco}, {Contessi},
  {Costa}, {Cutini}, {D'Ammando}, {Del Monte}, {de Paris}, {di Cocco}, {di
  Persio}, {Donnarumma}, {Evangelista}, {Fanari}, {Feroci}, {Ferrari},
  {Fiorini}, {Fornari}, {Fuschino}, {Froysland}, {Frutti}, {Galli},
  {Gasparrini}, {Labanti}, {Lapshov}, {Lazzarotto}, {Liello}, {Lipari},
  {Mattaini}, {Marisaldi}, {Mastropietro}, {Mauri}, {Mauri}, {Mereghetti},
  {Morelli}, {Moretti}, {Morselli}, {Pacciani}, {Perotti}, {Piano}, {Picozza},
  {Pilia}, {Pontoni}, {Porrovecchio}, {Preger}, {Prest}, {Primavera},
  {Pucella}, {Rapisarda}, {Rappoldi}, {Rossi}, {Rubini}, {Sabatini},
  {Santolamazza}, {Scalise}, {Soffitta}, {Stellato}, {Striani}, {Tamburelli},
  {Traci}, {Trois}, {Vallazza}, {Vittorini}, {Zambra}, {Zanello}, \&
  {Salotti}}]{2009A&A...506.1563P}
{Pittori}, C. {et~al.} 2009, \aap, 506, 1563

\bibitem[{{Raiteri} {et~al.}(2008{\natexlab{a}}){Raiteri}, {Villata}, {Chen},
  {Hsiao}, {Kurtanidze}, {Nilsson}, {Larionov}, {Gurwell}, {Agudo}, {Aller},
  {Aller}, {Angelakis}, {Arkharov}, {Bach}, {B{\"o}ttcher}, {Buemi},
  {Calcidese}, {Charlot}, {D'Ammando}, {Donnarumma}, {Forn{\'e}}, {Frasca},
  {Fuhrmann}, {G{\'o}mez}, {Hagen-Thorn}, {Jorstad}, {Kimeridze}, {Krichbaum},
  {L{\"a}hteenm{\"a}ki}, {Lanteri}, {Latev}, {Le Campion}, {Lee}, {Leto},
  {Lin}, {Marchili}, {Marilli}, {Marscher}, {Nesci}, {Nieppola},
  {Nikolashvili}, {Ohlert}, {Ovcharov}, {Principe}, {Pursimo}, {Ragozzine},
  {Sadun}, {Sigua}, {Smart}, {Strigachev}, {Takalo}, {Tavani}, {Thum},
  {Tornikoski}, {Trigilio}, {Uckert}, {Umana}, {Valcheva}, {Vercellone},
  {Volvach}, \& {Wiesemeyer}}]{rai08a}
{Raiteri}, C.~M. {et~al.} 2008{\natexlab{a}}, \aap, 485, L17

\bibitem[{{Raiteri} {et~al.}(2008{\natexlab{b}}){Raiteri}, {Villata},
  {Larionov}, {Gurwell}, {Chen}, {Kurtanidze}, {Aller}, {B{\"o}ttcher},
  {Calcidese}, {Hroch}, {L{\"a}hteenm{\"a}ki}, {Lee}, {Nilsson}, {Ohlert},
  {Papadakis}, {Agudo}, {Aller}, {Angelakis}, {Arkharov}, {Bach}, {Bachev},
  {Berdyugin}, {Buemi}, {Carosati}, {Charlot}, {Chatzopoulos}, {Forn{\'e}},
  {Frasca}, {Fuhrmann}, {G{\'o}mez}, {Gupta}, {Hagen-Thorn}, {Hsiao}, {Jordan},
  {Jorstad}, {Konstantinova}, {Kopatskaya}, {Krichbaum}, {Lanteri},
  {Larionova}, {Latev}, {Le Campion}, {Leto}, {Lin}, {Marchili}, {Marilli},
  {Marscher}, {McBreen}, {Mihov}, {Nesci}, {Nicastro}, {Nikolashvili}, {Novak},
  {Ovcharov}, {Pian}, {Principe}, {Pursimo}, {Ragozzine}, {Ros}, {Sadun},
  {Sagar}, {Semkov}, {Smart}, {Smith}, {Strigachev}, {Takalo}, {Tavani},
  {Tornikoski}, {Trigilio}, {Uckert}, {Umana}, {Valcheva}, {Vercellone},
  {Volvach}, \& {Wiesemeyer}}]{2008A&A...491..755R}
---. 2008{\natexlab{b}}, \aap, 491, 755

\bibitem[{{Raiteri} {et~al.}(2007){Raiteri}, {Villata}, {Larionov}, {Pursimo},
  {Ibrahimov}, {Nilsson}, {Aller}, {Kurtanidze}, {Foschini}, {Ohlert},
  {Papadakis}, {Sumitomo}, {Volvach}, {Aller}, {Arkharov}, {Bach}, {Berdyugin},
  {B{\"o}ttcher}, {Buemi}, {Calcidese}, {Charlot}, {Delgado S{\'a}nchez}, {di
  Paola}, {Djupvik}, {Dolci}, {Efimova}, {Fan}, {Forn{\'e}}, {Gomez}, {Gupta},
  {Hagen-Thorn}, {Hooks}, {Hovatta}, {Ishii}, {Kamada}, {Konstantinova},
  {Kopatskaya}, {Kovalev}, {Kovalev}, {L{\"a}hteenm{\"a}ki}, {Lanteri}, {Le
  Campion}, {Lee}, {Leto}, {Lin}, {Lindfors}, {Mingaliev}, {Mizoguchi},
  {Nicastro}, {Nikolashvili}, {Nishiyama}, {{\"O}stman}, {Ovcharov},
  {P{\"a}{\"a}kk{\"o}nen}, {Pasanen}, {Pian}, {Rector}, {Ros}, {Sadakane},
  {Selj}, {Semkov}, {Sharapov}, {Somero}, {Stanev}, {Strigachev}, {Takalo},
  {Tanaka}, {Tavani}, {Torniainen}, {Tornikoski}, {Trigilio}, {Umana},
  {Vercellone}, {Valcheva}, {Volvach}, \& {Yamanaka}}]{rai07}
---. 2007, \aap, 473, 819

\bibitem[{{Sakamoto} {et~al.}(2009){Sakamoto}, {D'Ammando}, {Gehrels},
  {Kovalev}, \& {Sokolovsky}}]{2009ATel.2329....1S}
{Sakamoto}, T., {D'Ammando}, F., {Gehrels}, N., {Kovalev}, Y.~Y., \&
  {Sokolovsky}, K. 2009, The Astronomer's Telegram, 2329, 1

\bibitem[{{Sasada} {et~al.}(2009){Sasada}, {Itoh}, {Yamanaka}, {Uemura},
  {Takahashi}, {Fukazawa}, {Kawabata}, {Ikejiri}, {Ohsugi}, \& {Kanata
  Team}}]{2009ATel.2333....1S}
{Sasada}, M. {et~al.} 2009, The Astronomer's Telegram, 2333, 1

\bibitem[{{Savolainen} {et~al.}(2009){Savolainen}, {Homan}, {Hovatta},
  {Kadler}, {Kovalev}, {Lister}, {Ros}, \& {Zensus}}]{2009arXiv0911.4924S}
{Savolainen}, T., {Homan}, D.~C., {Hovatta}, T., {Kadler}, M., {Kovalev},
  Y.~Y., {Lister}, M.~L., {Ros}, E., \& {Zensus}, J.~A. 2009, ArXiv e-prints

\bibitem[{{Striani} {et~al.}(2009{\natexlab{a}}){Striani}, {Vercellone},
  {Verrecchia}, {Pittori}, {Santolamazza}, {Tavani}, {D'Ammando}, {Donnarumma},
  {Vittorini}, {Del Monte}, {Evangelista}, {Feroci}, {Lazzarotto}, {Pacciani},
  {Soffitta}, {Costa}, {Lapshov}, {Rapisarda}, {Argan}, {Piano}, {Pucella},
  {Sabatini}, {Trois}, {Bulgarelli}, {Gianotti}, {Trifoglio}, {Fuschino},
  {Galli}, {Labanti}, {Marisaldi}, {di Cocco}, {Chen}, {Giuliani},
  {Mereghetti}, {Caraveo}, {Perotti}, {Pellizzoni}, {Pilia}, {Barbiellini},
  {Longo}, {Moretti}, {Vallazza}, {Morselli}, {Picozza}, {Prest}, {Lipari},
  {Zanello}, {Cattaneo}, {Rappoldi}, {Colafrancesco}, {Giommi}, \&
  {Salotti}}]{2009ATel.2322....1S}
{Striani}, E. {et~al.} 2009{\natexlab{a}}, The Astronomer's Telegram, 2322, 1

\bibitem[{{Striani} {et~al.}(2009{\natexlab{b}}){Striani}, {Vercellone},
  {Verrecchia}, {Pittori}, {Santolamazza}, {Tavani}, {D'Ammando}, {Donnarumma},
  {Vittorini}, {Del Monte}, {Evangelista}, {Feroci}, {Lazzarotto}, {Pacciani},
  {Soffitta}, {Costa}, {Lapshov}, {Rapisarda}, {Argan}, {Piano}, {Pucella},
  {Sabatini}, {Trois}, {Bulgarelli}, {Gianotti}, {Trifoglio}, {Fuschino},
  {Galli}, {Labanti}, {Marisaldi}, {di Cocco}, {Chen}, {Giuliani},
  {Mereghetti}, {Caraveo}, {Perotti}, {Pellizzoni}, {Pilia}, {Barbiellini},
  {Longo}, {Moretti}, {Vallazza}, {Morselli}, {Picozza}, {Prest}, {Lipari},
  {Zanello}, {Cattaneo}, {Rappoldi}, {Colafrancesco}, {Giommi}, \&
  {Salotti}}]{2009ATel.2326....1S}
---. 2009{\natexlab{b}}, The Astronomer's Telegram, 2326, 1

\bibitem[{{Tavani} {et~al.}(2009){Tavani}, {Barbiellini}, {Argan}, {Boffelli},
  {Bulgarelli}, {Caraveo}, {Cattaneo}, {Chen}, {Cocco}, {Costa}, {D'Ammando},
  {Del Monte}, {de Paris}, {di Cocco}, {di Persio}, {Donnarumma},
  {Evangelista}, {Feroci}, {Ferrari}, {Fiorini}, {Fornari}, {Fuschino},
  {Froysland}, {Frutti}, {Galli}, {Gianotti}, {Giuliani}, {Labanti}, {Lapshov},
  {Lazzarotto}, {Liello}, {Lipari}, {Longo}, {Mattaini}, {Marisaldi},
  {Mastropietro}, {Mauri}, {Mauri}, {Mereghetti}, {Morelli}, {Morselli},
  {Pacciani}, {Pellizzoni}, {Perotti}, {Piano}, {Picozza}, {Pontoni},
  {Porrovecchio}, {Prest}, {Pucella}, {Rapisarda}, {Rappoldi}, {Rossi},
  {Rubini}, {Soffitta}, {Traci}, {Trifoglio}, {Trois}, {Vallazza},
  {Vercellone}, {Vittorini}, {Zambra}, {Zanello}, {Pittori}, {Preger},
  {Santolamazza}, {Verrecchia}, {Giommi}, {Colafrancesco}, {Antonelli},
  {Cutini}, {Gasparrini}, {Stellato}, {Fanari}, {Primavera}, {Tamburelli},
  {Viola}, {Guarrera}, {Salotti}, {D'Amico}, {Marchetti}, {Crisconio},
  {Sabatini}, {Annoni}, {Alia}, {Longoni}, {Sanquerin}, {Battilana}, {Concari},
  {Dessimone}, {Grossi}, {Parise}, {Monzani}, {Artina}, {Pavesi},
  {Marseguerra}, {Nicolini}, {Scandelli}, {Soli}, {Vettorello}, {Zardetto},
  {Bonati}, {Maltecca}, {D'Alba}, {Patan{\'e}}, {Babini}, {Onorati},
  {Acquaroli}, {Angelucci}, {Morelli}, {Agostara}, {Cerone}, {Michetti},
  {Tempesta}, {D'Eramo}, {Rocca}, {Giannini}, {Borghi}, {Garavelli}, {Conte},
  {Balasini}, {Ferrario}, {Vanotti}, {Collavo}, \&
  {Giacomazzo}}]{2009A&A...502..995T}
{Tavani}, M. {et~al.} 2009, \aap, 502, 995

\bibitem[{{Tosti} {et~al.}(2008){Tosti}, {Chiang}, {Lott}, {Do Couto E Silva},
  {Grove}, \& {Thayer}}]{2008ATel.1628....1T}
{Tosti}, G., {Chiang}, J., {Lott}, B., {Do Couto E Silva}, E., {Grove}, J.~E.,
  \& {Thayer}, J.~G. 2008, The Astronomer's Telegram, 1628, 1

\bibitem[{{Vercellone} {et~al.}(2008){Vercellone}, {Chen}, {Giuliani},
  {Bulgarelli}, {Donnarumma}, {Lapshov}, {Tavani}, {Argan}, {Barbiellini},
  {Caraveo}, {Cocco}, {Costa}, {D'Ammando}, {Del Monte}, {De Paris}, {Di
  Cocco}, {Evangelista}, {Feroci}, {Fiorini}, {Froysland}, {Fuschino}, {Galli},
  {Gianotti}, {Labanti}, {Lazzarotto}, {Lipari}, {Longo}, {Marisaldi}, {Mauri},
  {Mereghetti}, {Morselli}, {Pacciani}, {Pellizzoni}, {Perotti}, {Picozza},
  {Prest}, {Pucella}, {Rapisarda}, {Soffitta}, {Trifoglio}, {Trois},
  {Vallazza}, {Vittorini}, {Zambra}, {Zanello}, {Pittori}, {Verrecchia},
  {Gasparrini}, {Cutini}, {Giommi}, {Antonelli}, {Colafrancesco}, \&
  {Salotti}}]{Vercellone2008:3C454_ApJ}
{Vercellone}, S. {et~al.} 2008, \apjl, 676, L13

\bibitem[{{Vercellone} {et~al.}(2009{\natexlab{a}}){Vercellone}, {Chen},
  {Vittorini}, {Giuliani}, {D'Ammando}, {Tavani}, {Donnarumma}, {Pucella},
  {Raiteri}, {Villata}, {Chen}, {Tosti}, {Impiombato}, {Romano}, {Belfiore},
  {DeLuca}, {Novara}, {Senziani}, {Bazzano}, {Fiocchi}, {Ubertini}, {Ferrari},
  {Argan}, {Barbiellini}, {Boffelli}, {Bulgarelli}, {Caraveo}, {Cattaneo},
  {Cocco}, {Costa}, {DelMonte}, {DeParis}, {Di Cocco}, {Evangelista}, {Feroci},
  {Fiorini}, {Fornari}, {Froysland}, {Fuschino}, {Galli}, {Gianotti},
  {Labanti}, {Lapshov}, {Lazzarotto}, {Lipari}, {Longo}, {Marisaldi},
  {Mereghetti}, {Morselli}, {Pellizzoni}, {Pacciani}, {Perotti}, {Picozza},
  {Prest}, {Rapisarda}, {Rappoldi}, {Soffitta}, {Trifoglio}, {Trois},
  {Vallazza}, {Zambra}, {Zanello}, {Pittori}, {Verrecchia}, {Santolamazza},
  {Preger}, {Gasparrini}, {Cutini}, {Giommi}, {Colafrancesco}, \&
  {Salotti}}]{Vercellone2008:3c454:ApJ_P1}
---. 2009{\natexlab{a}}, \apj, 690, 1018

\bibitem[{{Vercellone} {et~al.}(2010){Vercellone}, {D'Ammando}, {Vittorini},
  {Donnarumma}, {Pucella}, {Tavani}, {Ferrari}, {Raiteri}, {Villata}, {Romano},
  {Krimm}, {Tiengo}, {Chen}, {Giovannini}, {Venturi}, {Giroletti}, {Kovalev},
  {Sokolovsky}, \& {Pushkarev}}]{Vercellone2010:3C454}
---. 2010, \apj, 712, 405

\bibitem[{{Vercellone} {et~al.}(2009{\natexlab{b}}){Vercellone}, {Fiocchi},
  {Pian}, {Tavani}, {Bazzano}, {Ubertini}, {Argan}, {Costa}, {D'Ammando}, {Del
  Monte}, {Donnarumma}, {Evangelista}, {Feroci}, {Lapshov}, {Lazzarotto},
  {Pacciani}, {Piano}, {Pucella}, {Rapisarda}, {Sabatini}, {Soffitta},
  {Striani}, {Trois}, {Vittorini}, {Bulgarelli}, {Cocco}, {Fuschino}, {Galli},
  {Gianotti}, {Labanti}, {Marisaldi}, {Trifoglio}, {Caraveo}, {Chen},
  {Giuliani}, {Mereghetti}, {Perotti}, {Pellizzoni}, {Pilia}, {Barbiellini},
  {Longo}, {Moretti}, {Vallazza}, {Morselli}, {Picozza}, {Prest}, {Lipari},
  {Zanello}, {Cattaneo}, {Rappoldi}, {Colafrancesco}, {Giommi}, {Pittori},
  {Santolamazza}, {Verrecchia}, \& {Salotti}}]{2009ATel.2344....1V}
---. 2009{\natexlab{b}}, The Astronomer's Telegram, 2344, 1

\bibitem[{{Villata} {et~al.}(2007){Villata}, {Raiteri}, {Aller}, {Bach},
  {Ibrahimov}, {Kovalev}, {Kurtanidze}, {Larionov}, {Lee}, {Leto},
  {L{\"a}hteenm{\"a}ki}, {Nilsson}, {Pursimo}, {Ros}, {Sumitomo}, {Volvach},
  {Aller}, {Arai}, {Buemi}, {Coloma}, {Doroshenko}, {Efimov}, {Fuhrmann},
  {Hagen-Thorn}, {Kamada}, {Katsuura}, {Konstantinova}, {Kopatskaya}, {Kotaka},
  {Kovalev}, {Kurosaki}, {Lanteri}, {Larionova}, {Mingaliev}, {Mizoguchi},
  {Nakamura}, {Nikolashvili}, {Nishiyama}, {Sadakane}, {Sergeev}, {Sigua},
  {Sillanp{\"a}{\"a}}, {Smart}, {Takalo}, {Tanaka}, {Tornikoski}, {Trigilio},
  \& {Umana}}]{vil07}
{Villata}, M. {et~al.} 2007, \aap, 464, L5

\bibitem[{{Villata} {et~al.}(2006){Villata}, {Raiteri}, {Balonek}, {Aller},
  {Jorstad}, {Kurtanidze}, {Nicastro}, {Nilsson}, {Aller}, {Arai}, {Arkharov},
  {Bach}, {Ben{\'{\i}}tez}, {Berdyugin}, {Buemi}, {B{\"o}ttcher}, {Carosati},
  {Casas}, {Caulet}, {Chen}, {Chiang}, {Chou}, {Ciprini}, {Coloma}, {di Rico},
  {D{\'{\i}}az}, {Efimova}, {Forsyth}, {Frasca}, {Fuhrmann}, {Gadway}, {Gupta},
  {Hagen-Thorn}, {Harvey}, {Heidt}, {Hernandez-Toledo}, {Hroch}, {Hu}, {Hudec},
  {Ibrahimov}, {Imada}, {Kamata}, {Kato}, {Katsuura}, {Konstantinova},
  {Kopatskaya}, {Kotaka}, {Kovalev}, {Kovalev}, {Krichbaum}, {Kubota},
  {Kurosaki}, {Lanteri}, {Larionov}, {Larionova}, {Laurikainen}, {Lee}, {Leto},
  {L{\"a}hteenm{\"a}ki}, {L{\'o}pez-Cruz}, {Marilli}, {Marscher}, {McHardy},
  {Mondal}, {Mullan}, {Napoleone}, {Nikolashvili}, {Ohlert}, {Postnikov},
  {Pursimo}, {Ragni}, {Ros}, {Sadakane}, {Sadun}, {Savolainen}, {Sergeeva},
  {Sigua}, {Sillanp{\"a}{\"a}}, {Sixtova}, {Sumitomo}, {Takalo},
  {Ter{\"a}sranta}, {Tornikoski}, {Trigilio}, {Umana}, {Volvach}, {Voss}, \&
  {Wortel}}]{vil06}
---. 2006, \aap, 453, 817

\bibitem[{{Villata} {et~al.}(2009{\natexlab{a}}){Villata}, {Raiteri},
  {Gurwell}, {Larionov}, {Kurtanidze}, {Aller}, {L{\"a}hteenm{\"a}ki}, {Chen},
  {Nilsson}, {Agudo}, {Aller}, {Arkharov}, {Bach}, {Bachev}, {Beltrame},
  {Ben{\'{\i}}tez}, {Buemi}, {B{\"o}ttcher}, {Calcidese}, {Capezzali},
  {Carosati}, {da Rio}, {di Paola}, {Dolci}, {Dultzin}, {Forn{\'e}},
  {G{\'o}mez}, {Hagen-Thorn}, {Halkola}, {Heidt}, {Hiriart}, {Hovatta},
  {Hsiao}, {Jorstad}, {Kimeridze}, {Konstantinova}, {Kopatskaya}, {Koptelova},
  {Leto}, {Ligustri}, {Lindfors}, {Lopez}, {Marscher}, {Mommert}, {Mujica},
  {Nikolashvili}, {Palma}, {Pasanen}, {Roca-Sogorb}, {Ros}, {Roustazadeh},
  {Sadun}, {Saino}, {Sigua}, {Sorcia}, {Takalo}, {Tornikoski}, {Trigilio},
  {Turchetti}, \& {Umana}}]{2009A&A...504L...9V}
---. 2009{\natexlab{a}}, \aap, 504, L9

\bibitem[{{Villata} {et~al.}(2009{\natexlab{b}}){Villata}, {Raiteri},
  {Larionov}, {Konstantinova}, {Nilsson}, {Pasanen}, \&
  {Carosati}}]{2009ATel.2325....1V}
{Villata}, M., {Raiteri}, C.~M., {Larionov}, V.~M., {Konstantinova}, T.~S.,
  {Nilsson}, K., {Pasanen}, M., \& {Carosati}, D. 2009{\natexlab{b}}, The
  Astronomer's Telegram, 2325, 1

\bibitem[{{Woo} \& {Urry}(2002)}]{2002ApJ...579..530W}
{Woo}, J., \& {Urry}, C.~M. 2002, \apj, 579, 530

\end{thebibliography}

\end{document}